\documentclass[%
 aps,
 pre,%
 amsmath,amssymb,
reprint,%
superscriptaddress,showpacs]{revtex4-1}

\usepackage{color, graphicx}
\usepackage{dcolumn}
\usepackage{bm}
\usepackage{amssymb}
\usepackage{latexsym}
\usepackage{amsfonts}
\usepackage{amsmath}
\usepackage{multirow}
\usepackage{ifthen}

\begin{document}

\title{Energy transport in one-dimensional disordered granular solids}

\author{V.~Achilleos}
\author{G.~Theocharis}
\affiliation{LUNAM Universit\'e, Universit\'e du Maine, LAUM UMR CNRS 6613, Av. O. Messiaen, 72085 Le Mans, France}
\author{Ch.~Skokos}
\affiliation{Department of Mathematics and Applied Mathematics, University of Cape Town, Rondebosch 7701, South Africa
}

\begin{abstract}
We investigate the energy transport in one-dimensional disordered granular solids
by extensive numerical simulations. In particular, we consider the case of a polydisperse granular chain 
composed of spherical beads of the same material and with radii taken from 
a random distribution. 
We start by examining the linear case, in which it is known that the energy transport  strongly depends
on the type of initial conditions. Thus, we consider two sets of initial conditions: i) an initial displacement and ii) an initial momentum excitation of a single bead. 
After establishing the regime of sufficiently strong disorder, we focus our studies on the role of nonlinearity for both sets of initial conditions.
By increasing the initial excitation amplitudes we are able to identify three distinct dynamical regimes with different energy transport properties: a near linear, a weakly nonlinear and a highly nonlinear regime.
Although energy spreading is found to be increasing for higher nonlinearities, in the weakly nonlinear regime no clear asymptotic
behavior of the spreading is found. In this regime, we additionally find that energy, initially trapped in a localized region, can be eventually detrapped and this has a direct influence on the fluctuations 
of the energy spreading. We also demonstrate that in the highly nonlinear regime, the differences in energy transport between the two sets of initial conditions 
vanish. Actually, in this regime the energy is almost ballistically transported through shock-like excitations.

\end{abstract}

\maketitle

\section{Introduction}

{ Wave scattering and energy transport in disordered media have been for long time a matter of great research interest~\cite{books}.
The experimental observation of Anderson localization in different systems such as optical~\cite{andersonexpphot}, ultracold 
atomic gases~\cite{andersonexpBEC} and elastic networks~\cite{andersonexpselidas},
has renewed the research in this direction.  In addition, recent studies on wave scattering in random media have lead to a plethora of applications in imaging, random lasing and solar energy (see for example~\cite{wiersma} and references therein).

The key phenomenon employing in these studies is the spatial wave localization due to disorder, which is a linear effect relying on 
keeping phase coherence of participating waves~\cite{Anderson}. However, wave localization can also emerge due to 
nonlinearity, as it was first shown in the studies of Fermi-Pasta-Ulam (FPU)~\cite{FPU}, and may lead to energy localization and propagation through the formation of localized solutions (solitons, breathers etc.) in different  lattice models~\cite{panos}.
The interplay of these two localization mechanisms, nonlinearity and disorder}, has
been studied extensively in the recent years~\cite{dn1,dn2,dn3,dn4,dn5,dn6,dn7,dn8,dn9,dn10,dn11,dn12,dn13,dn14,dn15,Bourbonnais,Zavt,lepriPRE}.
In most of these studies,  an initially localized wavepacket was shown to lead to delocalization and a sub-diffusive spreading of the energy, for sufficiently large nonlinearities.
The most common models that have been studied are the Klein-Gordon (KG) model as well as 
the discrete nonlinear Schr\"{o}dinger (DNLS),
where especially the latter has attracted much attention due to its application
to various optical structures and devices. Experimental studies on optical structures show that nonlinearity
can either enhance localization (for focusing nonlinearity) or induce delocalization 
(for defocusing nonlinearity)~\cite{dnlp1,dnlp2}.

{ Granular solids, namely densely packed arrays of macroscopic particles which appear naturally disordered, are a promising testbed for studying the interplay of disorder and nonlinearity. The latter originates from the interparticle Hertzian contacts~\cite{hertzbook}. An especially appealing characteristic of these media is their tunable dynamical response} ranging from near linear to highly nonlinear, 
by changing the ratio of static to dynamic interparticle displacements. 
{ Fabricated granular solids} have allowed the exploration of a plethora of 
fundamental phenomena, including { solitary waves with a highly localized waveform} in the case of uncompressed
crystal, discrete breathers and others ~\cite{compacton,gr1,gr2,gr3,gr4,gr5,gr6,chapterG}. They have been also applied in various engineering devices, including among others shock and energy absorbing 
layers~\cite{dar06,hong05,doney06}, acoustic lenses \cite{Spadoni}, and acoustic 
diodes \cite{Nature11}. 

For sufficiently weak excitations and in the presence of precompression, the one-dimensional disordered granular solid, also called granular chain, can be approximated by a disorder harmonic lattice which has some interesting transport properties.
In particular, it has been shown that different initial conditions
-- initial displacement or momentum excitations -- of a single particle can lead to: sub-diffusive (displacement) or super-diffusive (momentum) energy transport~\cite{kundu}, as well as to analytical described asymptotic energy profiles~\cite{lepriPRE}.
{On the other hand, for sufficiently strong excitations or in the absence of precompression, the granular chain exhibits two different types of nonlinearity:}
(i) a power nonlinearity stemming from the Hertzian contacts and (ii) a non-smooth nonlinearity, which is triggered whenever two beads of the chain lose contact (gap opening).
{ The latter is present to a broad class of fragile mechanical systems that loose rigidity upon lowering the external pressure towards zero, such as weakly connected polymers~\cite{poly} and network glasses~\cite{glasses}. It is also present in cracked solids~\cite{cracked}.}

{ Recently, studies of one-dimensional 
disordered granular chains
have been reported~\cite{Luding,geubelle1,chiaropanos}.
In the absence of precompression, where the {\it non-smooth nonlinearity} is present, it was shown that if a solitary wave is formed, it features an exponential decay which strongly depends
on the degree of randomness~\cite{geubelle1,chiaropanos}.}
{ Similar results were also reported in a two-dimensional granular solid~\cite{vitelli1}} where the decay of the amplitude of the wave front was
described using an analogy between disorder and viscoelastic dissipation.
On the other hand, { in the presence of precompression}, the power nonlinearity stemming from the Hertzian contacts leads to a FPU like dynamics, which
have been studied theoretically in the presence of disorder~\cite{lepriPRE,Bourbonnais,Zavt}. However, in the case of granular chains, { this dynamics} can be strongly modified by the presence of the opening of
gaps. Thus, the interplay of these two nonlinear mechanisms is of particular interest and it can drastically change the transport properties. 

{ Only recently, a study about one-dimensional and precompressed random dimer granular chains~\cite{MasonPanos} has reported some features of the energy transport. In this work, the authors compare wave dynamics in chains with three different types of disorder: an uncorrelated (Anderson-like) disorder and two types of correlated disorder. For the Anderson-like uncorrelated disorder, they
found a transition from subdiffusive to superdiffusive dynamics depending on the amount of precompression in the chain. In the present work, we consider  a different kind of uncorrelated disorder (i.e. polydispersity through disorder in the bead radius) and we study both displacement and momentum initial excitations, emphasizing their differences and similarities~\cite{Mason2}.
In particular, we consider polydisperse disordered granular chains composed of spherical beads of the same material and with radii taken from a random distribution. Our motivation is the fact that most of the granular materials occurring in nature and industrial application are composed of a broad range of particle sizes~\cite{poly}.} By considering a single central bead excitation, we study the transport of energy in these disordered granular chains.

 In Sec.~\ref{s1} we present the equations of motion in a normalized
form, we define the conserved energy of the system and also describe the parameters used for the characterization of the energy transport.
Results for the linear case are shown in Sec.~\ref{s2} where the influence of the strength of the disorder on the dynamics is studied. 
In Sec.~\ref{s3} we show the 
main results of this work for the case of an initial displacement excitation, and explicitly identify three different regions of energy
 transport, a near linear one, a weakly nonlinear and a highly nonlinear regime. The energy transport for the case of an initial 
 momentum excitation is discussed in  Sec.~\ref{s4}, presenting differences and similarities with the initial displacement excitation.
Results  concerning the asymptotic profile of the energy for the two types of initial conditions are shown in Sec.~\ref{s5}
and finally in Sec.~\ref{s6} we conclude our results.

\section{Disordered granular chain}
\label{s1}
\begin{figure}
\includegraphics[scale=0.22]{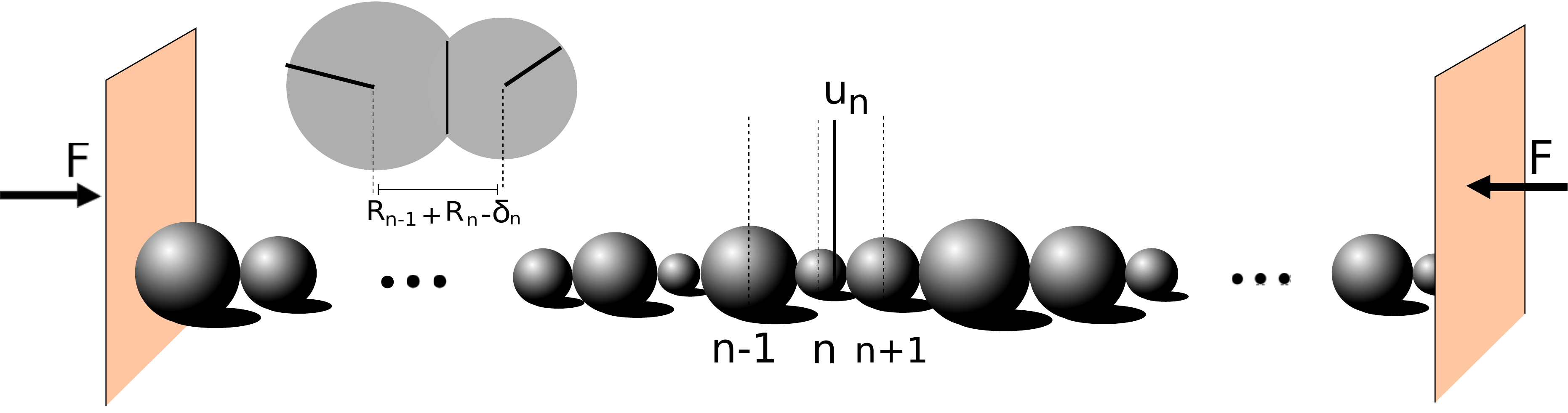}
\caption{A sketch of a  granular chain with beads of random radius. $u_n$ denotes the displacement of each bead 
from its equilibrium position, while $\delta_n$ is the overlap between two spherical beads due to the precompression force $F$.}
\label{sketch}
\end{figure}

We consider a one-dimensional chain consisting of $N+2$ spherical beads,
with masses $m_n$ ($n=0,1,2\ldots N+1$) and Hertzian contacts as shown in Fig.~\ref{sketch}. 
{ We consider fixed boundary conditions for the first and last spherical bead, namely~$u_0=u_{N+1}=0$,
where $u_n$ is the displacement of each bead from its equilibrium position. Then,} the system is described by the following 
set of differential equations:
\begin{align}
&m_1\ddot{u}_1=A_{1}[\delta_1-u_1]_+^{\frac{3}{2}}-A_{2}[\delta_{2}+u_1-u_2]_+^{\frac{3}{2}},\nonumber \\
&m_n\ddot{u}_n=A_{n}[\delta_n\!+\!u_{n-1}-u_n]_+^{\frac{3}{2}}-A_{n+1}[\delta_{n+1}\!+\!u_n-u_{n+1}]_+^{\frac{3}{2}},
\nonumber	\\
&m_N\ddot{u}_N=A_{N}[\delta_{N}+u_{N-1}-u_N]_+^{\frac{3}{2}}-A_{N+1}[\delta_{N+1}+u_{N}]_+^{\frac{3}{2}},
\label{nlineareqm}
\end{align}
where 
$A_n$ is the contact coefficient between beads $n-1, n$ 
and $\delta_{n}$ is the relative static overlap due to a precompression force $F$ acting on the two boundaries.
 The dots ($\dot{ }$)
denote differentiation with respect to time.
The coefficient $A_n$ for spherical beads of the same material, is given 
by $A_n=(2/3)\mathcal{E}\sqrt{(R_{n-1}R_{n})/(R_{n-1}+R_n)}/(1-\nu^2)$~\cite{hertzbook},
where $\mathcal{E}$ and $\nu$  are the elastic modulus and Poisson's ratio respectively, while $R_n$ is the radius of the $n$th bead.
The static overlap $\delta_n$ is given by $\delta_n=(F/A_n)^{2/3}$~\cite{hertzbook}.
The $[\quad]_+$ sign in Eq.~(\ref{nlineareqm}) denotes the following: when the expression inside the square 
brackets becomes negative (i.e the beads are not in contact) this term becomes zero. In fact this happens when the 
relative displacement between two beads becomes larger than their overlap  $u_{n-1}-u_n>\delta_n$, that is  
there is a \textit{gap} between them, and their relative force vanishes.

{ Below we will work in dimensionless units, however 
for clarity we note that we use a reference  radius of $R=0.01$~m, a static force of $F=1$~N and stainless 
steel spherical beads (316 type), the elastic modulus of which is $\mathcal{E}=193$ GPa while the Poisson ratio is $\nu= 0.3$. Relevant experiments with granular chains contained few defects can be found in~\cite{def_exps}.} In the following we will consider a disordered setup where the radii $R_n$ of the different beads
will be taken as a random variable, with values taken from a uniform distribution within the range
$R_n\in [R,\alpha R]$, where the parameter $\alpha\ge 1$ describes the disorder strength. Consequently, the  
mean value  $\tilde{R}$ of the bead radius is $\tilde{R}=(\alpha+1)R/2$. 
In order to make our equations dimensionless we implement the following transformations for time,
distance, mass and stiffness respectively:
\begin{equation}
\begin{aligned}
&t\rightarrow \tilde{\omega}t,\;\; \delta_n\rightarrow\delta_n/\tilde{\delta}\; (u_n\rightarrow u_n/\tilde{\delta}),\;\;
 \\
&m_n\rightarrow m_n/\tilde{m},\;\; A_n\rightarrow A_n/6\tilde{A}, \label{norm}
\end{aligned}
\end{equation}
where all the quantities with tilde, are calculated at $\tilde{R}$.
The frequency  $\tilde{\omega}_c=(6\tilde{A}\tilde{\delta}^{1/2}/\tilde{m})^{1/2}$ is the
cutoff frequency corresponding to the linear case of a chain with spherical beads of radius $\tilde{R}$.
The normalization is such that in the case of no disorder ($\alpha=1$) the \textit{normalized} cutoff frequency is $\omega_c=1$.
%
The energy of the system is given by the following expression:
\begin{eqnarray}
E=\sum_{n=1}^{N} E_n\equiv \sum_{n=1}^{N}\left( \frac{p_n^2}{2m_n} +V_{n}\right),
\label{Hamlinear}
\end{eqnarray}
where $E_n$ and $p_n=m_n\dot{u}_n$ are the energy and momentum of the $n$th bead respectively.
The potential $V_n$ for each spherical bead is defined as $V_n=[V(u_{n})+V(u_{n+1})]/2$ where:
\begin{equation}
\begin{aligned}
&V(u_n)=\frac{2}{5}A_n[\delta_n+u_{n-1}-u_n]_+^{5/2}-\frac{2}{5}A_n\delta_n^{5/2} \\
&-A_n\delta_n^{3/2}(u_{n-1}-u_n).
\end{aligned}
\label{Vhz}
\end{equation}

{ To study the energy transport in this one-dimensional system we focus on the time evolution of
the second moment of the energy distribution~\cite{lepriPRE} defined as:}
\begin{eqnarray}
m_2=\frac{\sum_n|n-n_0|^2 h_n(t)}{E},
\label{moments}
\end{eqnarray}
where $n_0=N/2$ corresponds to the central bead of the chain and $N=2\times 10^4$ to the total number of the spherical beads.
In the last expression, $h_n=E_n/E$ denotes the portion of the total energy $E$ acquired by the
{ $n$th} bead. Another useful quantity that characterizes the system is the participation number:
\begin{eqnarray}
P=1/\sum_n h_n^2,
\end{eqnarray}
{ which measures the number of excited beads that significantly contribute in the energy distribution}.
It takes the value  $P=1$ if all the energy is
concentrated in one bead while it becomes $P=N$ in the case of energy equipartition.
In our study, we investigate the energy transport under two different 
sets of initial conditions:
\begin{itemize}
\item  $u_{N/2}(0)=u,\;\; u_{n\ne N/2}(0)=0,\;\;\dot{u}_n(0)=0,$
\item  $u_n(0)=0,\;\; \dot{u}_{N/2}(0)=\dot{u},\;\;\dot{u}_{n\ne N/2}(0)=0,$
\end{itemize}
corresponding respectively  to an initial \textit{displacement} and an initial \textit{momentum} excitation of the central bead. 
We present  results obtained by averaging over  $200$ disorder realizations.
Throughout the text the average value over disorder realizations of a quantity $x$, is denoted as $\langle x \rangle$.
Simulations are carried out using the SABA2C symplectic integrator which allows as to keep  the relative energy error
at the order of $10^{-4}$~\cite{saba2,dn5}.  We also note that in our simulations energy never reaches the boundaries of the chain.
\begin{figure}
\includegraphics[scale=0.36]{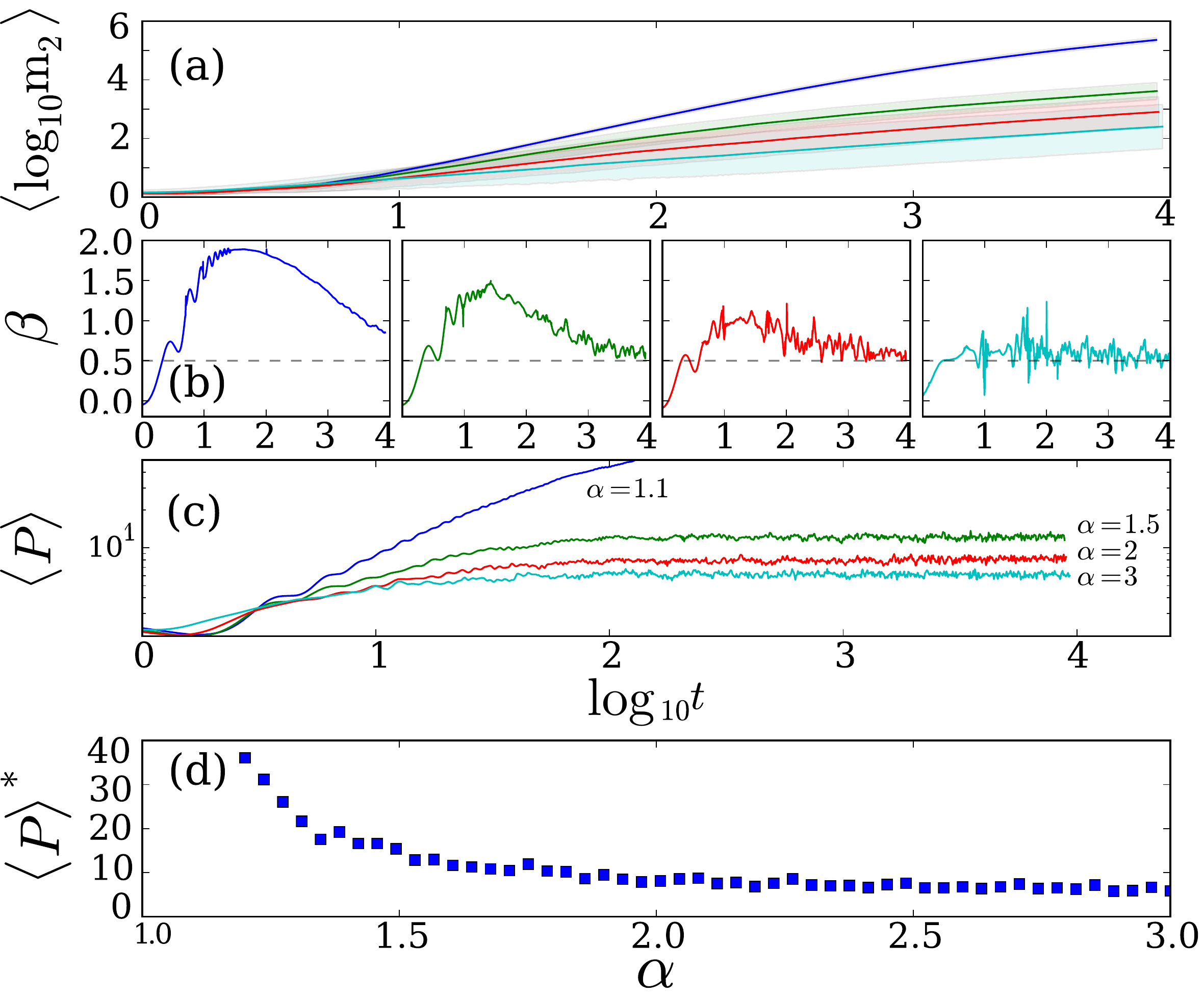}
\caption{(Color online)~{Harmonic chain: (a) Time evolution of $\langle \log_{10}m_2(t)\rangle$ for different values of $\alpha$.
The shaded area corresponds to the standard mean deviation of the measured mean value. Curves from top to bottom
correspond to $\alpha=1.1, 1.5, 2, 3$ respectively.
(b) The time derivative $\beta$ of $\langle \log_{10}m_2(t)\rangle$  given by Eq.~(\ref{beta}),
as a function of  $\log_{10}t$ for $\alpha=1.1, 1.5, 2, 3$ from left to right. The horizontal dashed lines indicate the value $\beta=0.5$.
(c)  Time evolution of the mean participation number $\langle P\rangle$ for the different values 
of the disorder parameter $\alpha$. Curves from top to bottom
correspond to $\alpha=1.1, 1.5, 2, 3$ respectively. (d) The asymptotic value of the mean participation number $\langle P\rangle ^*$ as a function of $\alpha$.}}
\label{linplot}
\end{figure}

\section{Harmonic chain}
\label{s2}
For sufficiently small displacements, i.e. { $u_{n-1}-u_n\ll \delta_n$}, 
the system of Eqs.~(\ref{nlineareqm}) can be approximated by the following linear system:
\begin{align}
m_n\ddot{u}_n=K_n (u_{n-1}-u_n)-K_{n+1}(u_n-u_{n+1}), \label{lineareqm} 
\end{align}
where $K_n=(3/2)A_n\delta_n^{1/2}$ is the linear coupling constant.
In the absence of disorder, $\alpha=1$, the energy spreading is ballistic and the second moment 
grows in time as $m_2(t)\sim t^2$, while $\langle P\rangle$ diverges for both displacement  
and momentum initial excitations. 
On the other hand, for the case of randomly chosen radii, Eq.~(\ref{lineareqm}) has the form of a disordered harmonic chain. 
This system has already been studied in several works~\cite{ishi,kundu,lepriPRE} with either a mass disorder 
or a disorder in the coupling constants $K_n$. 
For  the  granular chain considered here, having beads of the same material but of different radius,
both the  masses $m_n$ and the  coupling constants $K_n$ are random variables (both depend on the radius of the beads). 
Since masses depend on the radius as { $m\propto R^3$}, while the linear couplings as {$K\propto R^{1/3}$},we expect that the disorder effect is stronger due to the masses. 

In order to investigate the importance of the disorder parameter $\alpha$ on the system's behavior, we numerically
integrate Eqs.~(\ref{lineareqm})  for different values of $\alpha$ and the results are presented in Fig.~\ref{linplot}.
In particular, in Fig.~\ref{linplot} (a),  we show the time evolution of the average logarithm of the second moment $\langle \log_{10}m_2(t)\rangle$ 
with respect to the logarithm of time.  Furthermore in the panels of Fig.~\ref{linplot}  (b),  we show the time evolution of $\beta$, the time derivative of $\langle \log_{10}m_2(t)\rangle$  given by
\begin{eqnarray}
\beta=\frac{d\langle \log_{10}m_2\rangle}{d \log_{10}t}.
\label{beta}
\end{eqnarray}
The derivative is calculated numerically as follows: we first smooth the values of $\langle \log_{10}(m_2)\rangle$ by using a locally
weighted regression algorithm~\cite{clev}, and then we apply an $8^{th}$ order central finite difference scheme to
compute the derivative.
In~\cite{kundu,lepriPRE}, it was shown that for an initial displacement the energy transport is sub-diffusive. In particular,
the second moment grows in time as $m_2(t)\sim t^\beta$ with  an asymptotic value for the  exponent  $\beta(t\rightarrow\infty)\sim 0.5$. From the left panel
of  Fig.~\ref{linplot} (b), its readily seen that for $\alpha=1.1$ the parameter $\beta(t)$ initially acquires a value close to $\beta=2$
indicating a ballistic spreading of energy, but eventually it drops to smaller values implying a slower spreading. However,
one can not induce a clear sub-diffusive behavior for the time scales of our simulations. 
The second and third panels of Fig.~\ref{linplot} (b) correspond to stronger disorder ($\alpha=1.5$ and $\alpha=2$ respectively) 
where { the tendency of $\beta$ to asymptotically reach the value of $\beta=0.5$ is evident}, although some larger fluctuations are present. 
In the rightmost panel of Fig.~\ref{linplot} (b), we plot the case of $\alpha=3$, we see that the value of $\beta$ saturates
to $\beta=0.5$ very fast. However, we also notice that for this value of $\alpha$, the fluctuations are even larger and this is also depicted by the very large standard deviation in the mean value of  $\langle \log_{10}m_2(t)\rangle$,  indicated by the shaded area around  the lowest curve of Fig.~\ref{linplot} (a). 
Therefore, a much larger number of realizations is needed for a better  analysis of the $\alpha=3$ case.

The respective mean participation  number $\langle P\rangle $ for the different values of the disorder parameter $\alpha$ is
shown in Fig.~\ref{linplot} (c). { For  $\alpha=1.1$, the mean participation number does not saturate, at least on
the times of our simulations. On the other hand, for stronger disorder e.g. $\alpha=2$, it acquires an asymptotic value depending 
on the disorder parameter. The dependence of the numerical estimation of the asymptotic value of 
$\langle P(t\rightarrow \infty)\rangle=\langle P\rangle^*$ for different values of $\alpha$ is shown in Fig.~\ref{linplot} (d). 
These estimations are obtained using the results of Fig.~\ref{linplot} (c) as the mean value of $\langle P\rangle$ at the second half
of the last decade of the simulation i.e. for $5\times 10^3\leq t \leq10^4$.
Thus we may identify: a weak disorder regime ($\alpha\lesssim 1.5$) where $\langle P\rangle^*$ 
has a value larger than 10 and approaches the total number of particles $N$ in the case of no disorder ($\alpha=1$), and a strong disorder regime  ($\alpha>1.5$) where it saturates to values of  about 10 beads or less.
For this reason and due to the fact that for values $\alpha\ge 3$ the smoothing of large fluctuations of the computed quantities would imply many more disorder realizations, we restrict our analysis to 
the disorder regime with $\alpha=2$. }

We note that similar results to the ones of Fig.~\ref{linplot} were obtained  for the case of initial momentum excitation. In particular, we recovered  that the asymptotic value of $\beta$ is 1.5,  in the case of $\alpha=2$.
In the rest of this work we systematically investigate the effect of the nonlinearity to our system.
\begin{figure}
\includegraphics[scale=0.38]{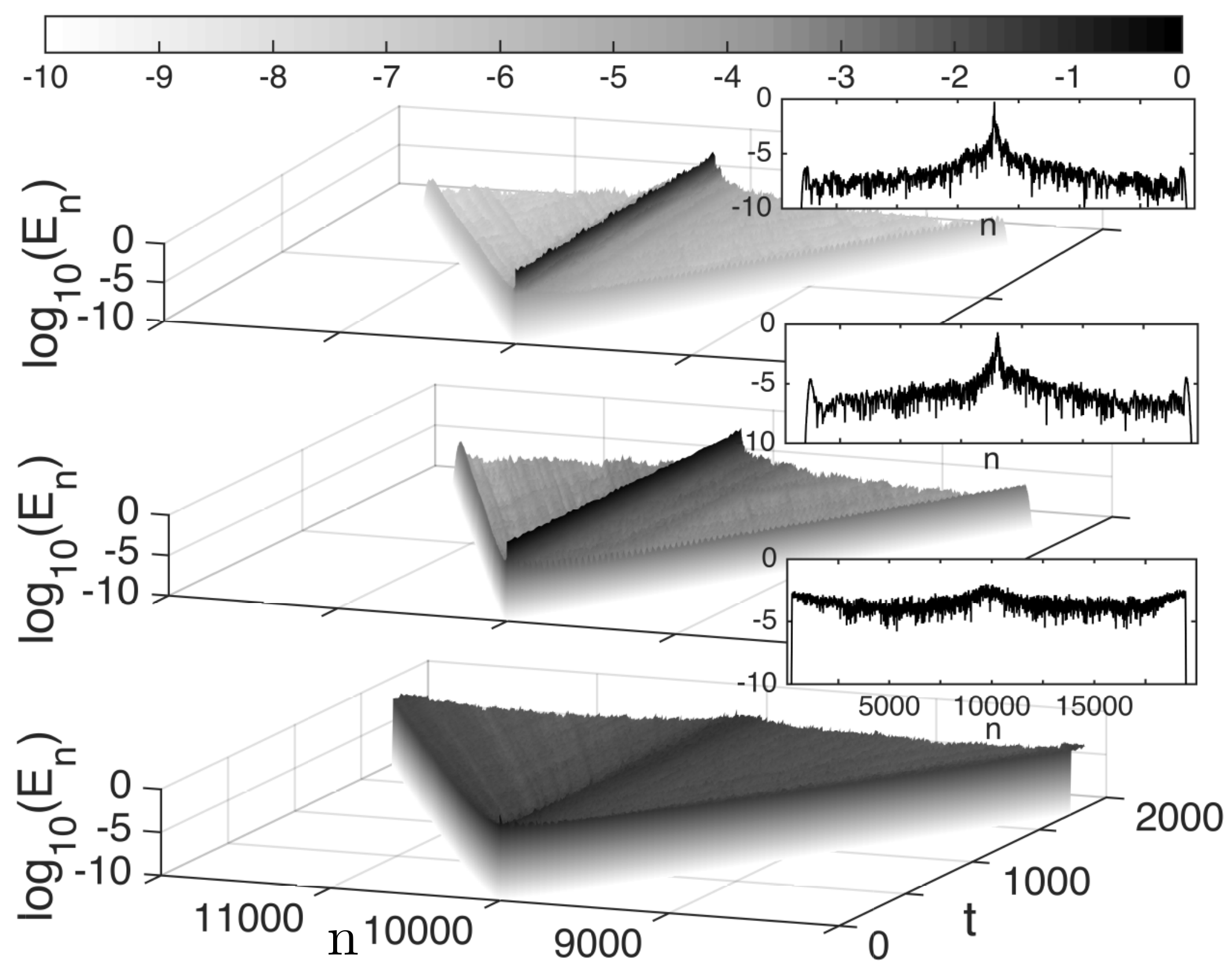}
\caption{Time evolution of the logarithm of the energy $E_n$ of a portion of the chain near the initially 
excited bead at $n=N/2=10^4$. Different panels (from top to bottom) correspond to { different values of the initial displacement $u$, and in particular to $u=0.01$, $1$, and $10$ respectively}. 
The coloring of each lattice site, according to the color scale shown on top of the panels, denotes the $\log_{10}(E_n)$ value of the corresponding bead. 
The insets on the right of each panel show the corresponding energy profile at a late time instant, $t=2\times10^3$.}
\label{3d}
\end{figure}

\section{Displacement excitation}
\label{s3}

In this Section we  present our findings for the transport of energy induced by an initial displacement of the central spherical bead in the chain.
Typical results of the dynamics observed for different amplitudes of the initial displacements, are shown in Fig.~\ref{3d}.
The top and middle panels illustrate clearly that a localized state is formed near the initially excited bead, while there are two
propagating fronts traveling towards the edges of the chain. These results show that for an increasing amplitude of the initial
excitation, these fronts are found to be propagating faster. In the bottom panel, which corresponds to $u=10$,
the behavior is significantly altered, since now  the energy looks to be more equally distributed through the lattice.
This fact is better illustrated by looking 
at the insets (in the right of each panel), where we show the corresponding energy profile at a late time instant $t=2\times10^3$ for each case. 
For the cases of the top and middle panels, it is readily seen that the energy around the initially excited bead is almost five orders of magnitude larger compared to the energy of more distant beads. 
{ On the other hand, in the bottom panel the differences of energy between the central region and the rest of the chain are 
significantly smaller.}

In what follows we discuss in more detail the outcomes of extensive numerical simulations for several values of the 
nonlinearity strength. The main results are summarized in Fig.~\ref{singsite}. In particular, in Fig.~\ref{singsite} (a)
we plot the time evolution of $\langle \log_{10} m_2\rangle$, in Fig.~\ref{singsite} (b) we plot the time evolution of its 
time derivative and finally in  Fig.~\ref{singsite} (c)  we show the mean  participation number $\langle P\rangle$
as a function of time.

\subsection{Near linear regime}

\begin{figure}
\includegraphics[scale=0.36]{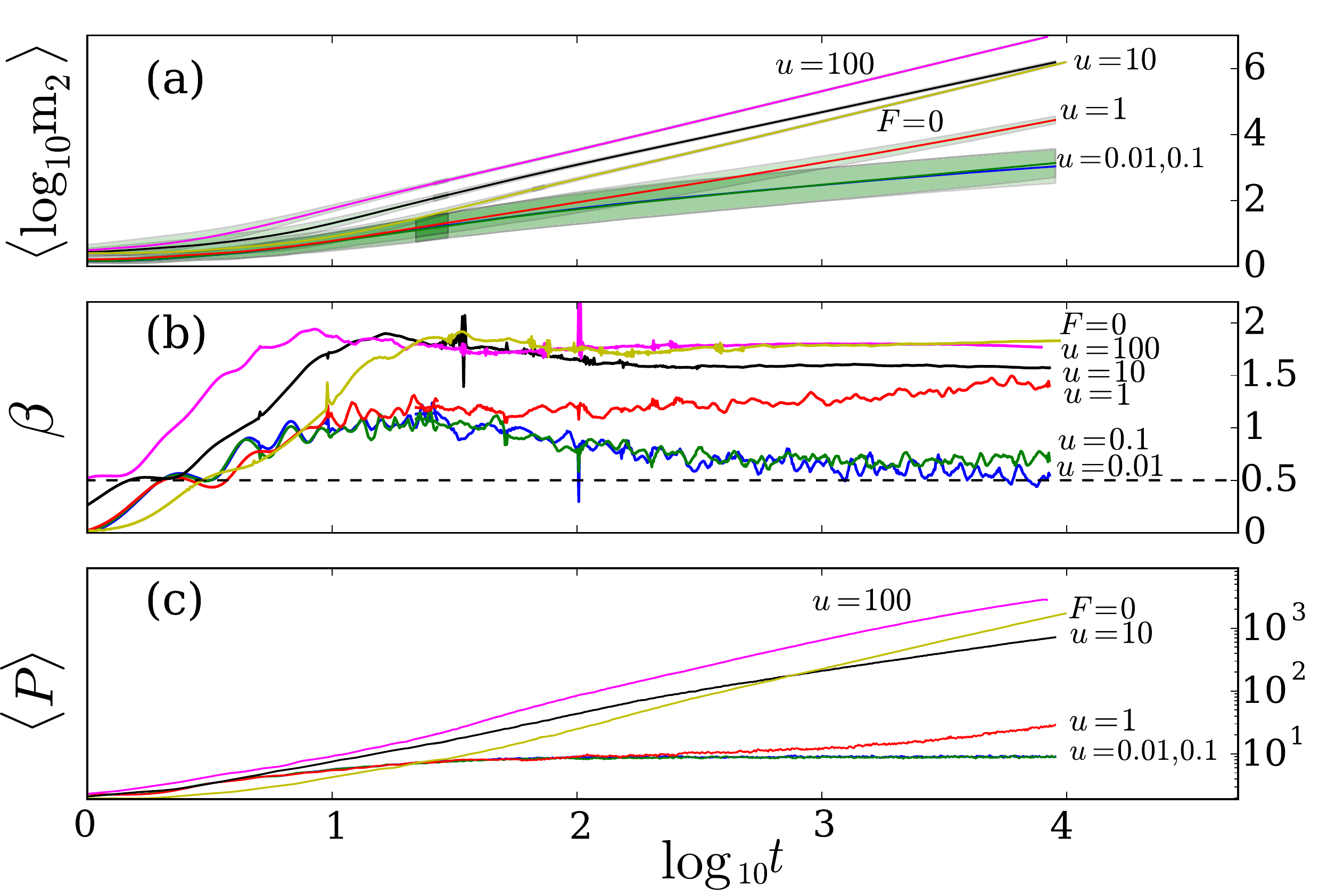}
\caption{(Color online) Time evolution of: (a) the averaged logarithm of the second moment $\langle \log_{10}m_2\rangle$ 
for different initial displacements $u$, (b) { the time derivative $\beta(t)$ of the respective curves of panel (a)}, (c) the mean value of the participation number 
$\langle P\rangle$.
}
\label{singsite}
\end{figure}

Let us first note that for values of the initial excitation  $u<0.1$ 
[see e.g.~the blue (lowest) line in Fig.~\ref{singsite} (b) which corresponds to $u=0.01$], we observe that  {$\beta\approx 0.5$} at about {$t\approx 10^4$}.
Additionally the mean participation  number for this case [Fig.~\ref{singsite} (c)] is found to practically saturate 
to the same value { as in the harmonic chain.}  Thus, we conclude that for values
of $u<0.1$ the nonlinear model is characterized by a \textit{near linear} regime, having qualitatively 
and quantitatively the same energy transport properties as its linear counterpart, at least for the time scales of our simulations.

\subsection{Weakly nonlinear regime}

From the results of Fig.~\ref{singsite} it is readily seen that already 
for an  initial condition of $u=0.1$,  the value of $\beta$ deviates from its behavior in the linear case, { becoming larger than $0.5$.} 
However a very abrupt change in the behavior of $\beta$ is  observed at the value $u=1$ 
and the system undergoes a transition from sub-  to super-diffusion. Notice also that although for $u=0.1$ the 
mean participation number in Fig.~\ref{singsite} (c),  saturates to a value of $\langle P\rangle \approx 10$, for the  value
$u=1$ it is found to { continuously increase}. These two results indicate the existence of a regime of intermediate dynamics 
for values between $u=0.1$ to $u=1$.

\begin{figure}
\includegraphics[scale=0.37]{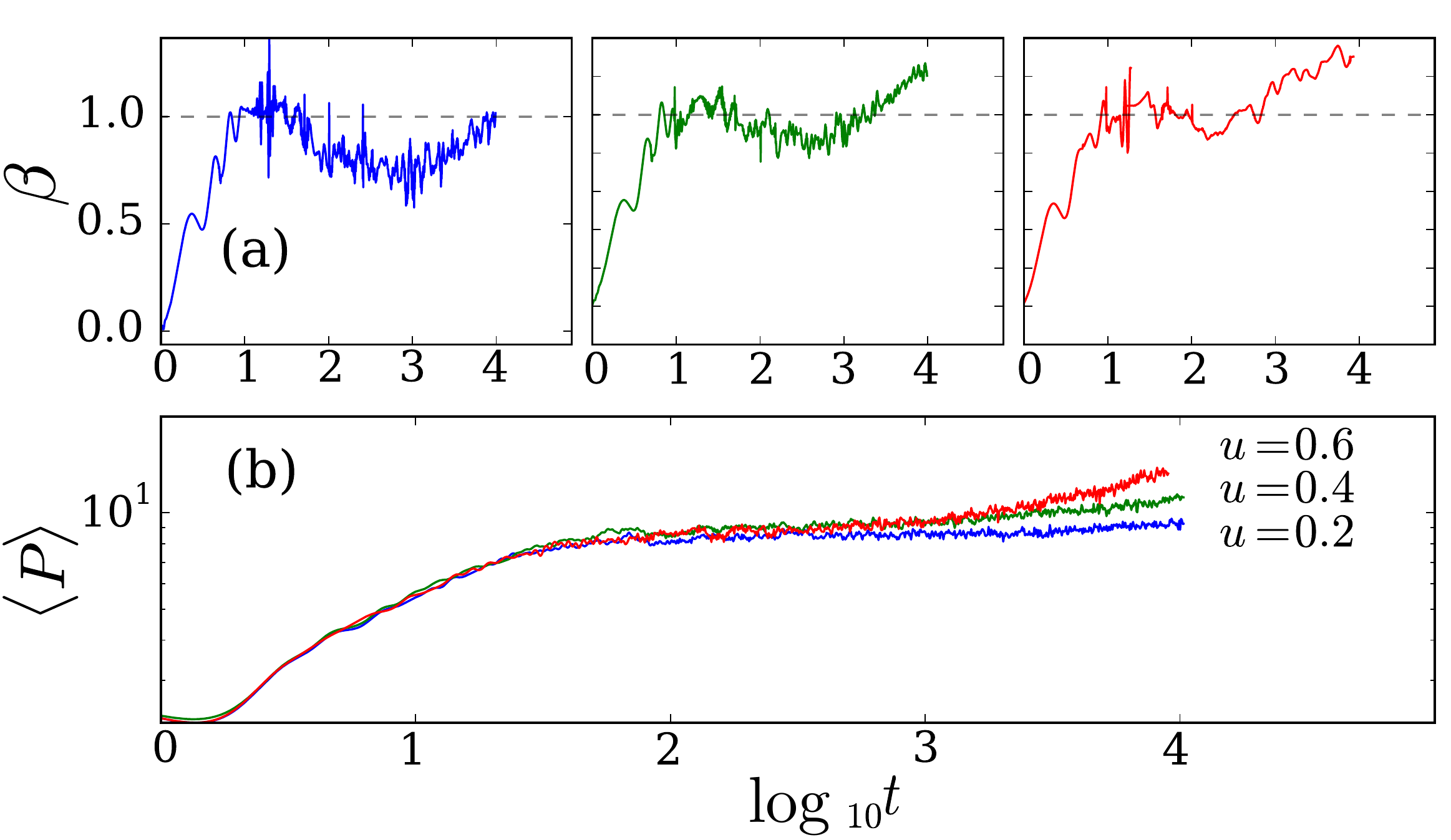}
\caption{{ (Color online) (a): Time evolution of the parameter $\beta$ for different values of the initial displacement excitation { 
$u=0.2, 0.4, 0,6$ (from left to right),} in the regime where energy transport crosses from sub- to super-diffusive behavior. The horizontal dashed lines indicate the value $\beta=1$. (b) Time evolution of the mean participation number for displacement excitations
corresponding to the top panels.}}
\label{intermediate}
\end{figure}

In order to further understand the dynamics in this regime, we compute the parameter $\beta$,
as well as the mean participation number $\langle P\rangle$, for some intermediate values of initial displacements  i.e.~$u=0.2, 0.4, 0.6$. 
The obtained  results, which are plotted in Fig.~\ref{intermediate}, clearly show that a transition from sub- to super-diffusion is 
carried out in this regime. In Fig.~\ref{intermediate} (a), the parameter $\beta$ exhibits many fluctuations and shows no evident tendency to 
saturate into a constant value until the end of our numerical simulations. In all cases 
shown in Fig.~\ref{intermediate}(a),  $\beta$ initially approaches the diffusive value  $\beta=1$ but later 
on starts to decrease.  Furthermore at a time interval between 
{ $t\approx10^2$ and $t\approx10^3$} it saturates to an almost constant value somewhat below {$\beta=1$}, but eventually the dynamics 
changes and $\beta$ starts to increase again becoming larger than $1$. 
This behavior creates a characteristic local minimum of $\beta(t)$ for all studied cases shown in Fig.~\ref{intermediate}. 
This result is in accordance with the recent study of~\cite{lepriPRE}, where it was found that in the FPU problem,
until the end of the studied times, there was no clear asymptotic value for the exponent of $m_2$.

\begin{figure}
\includegraphics[scale=0.37]{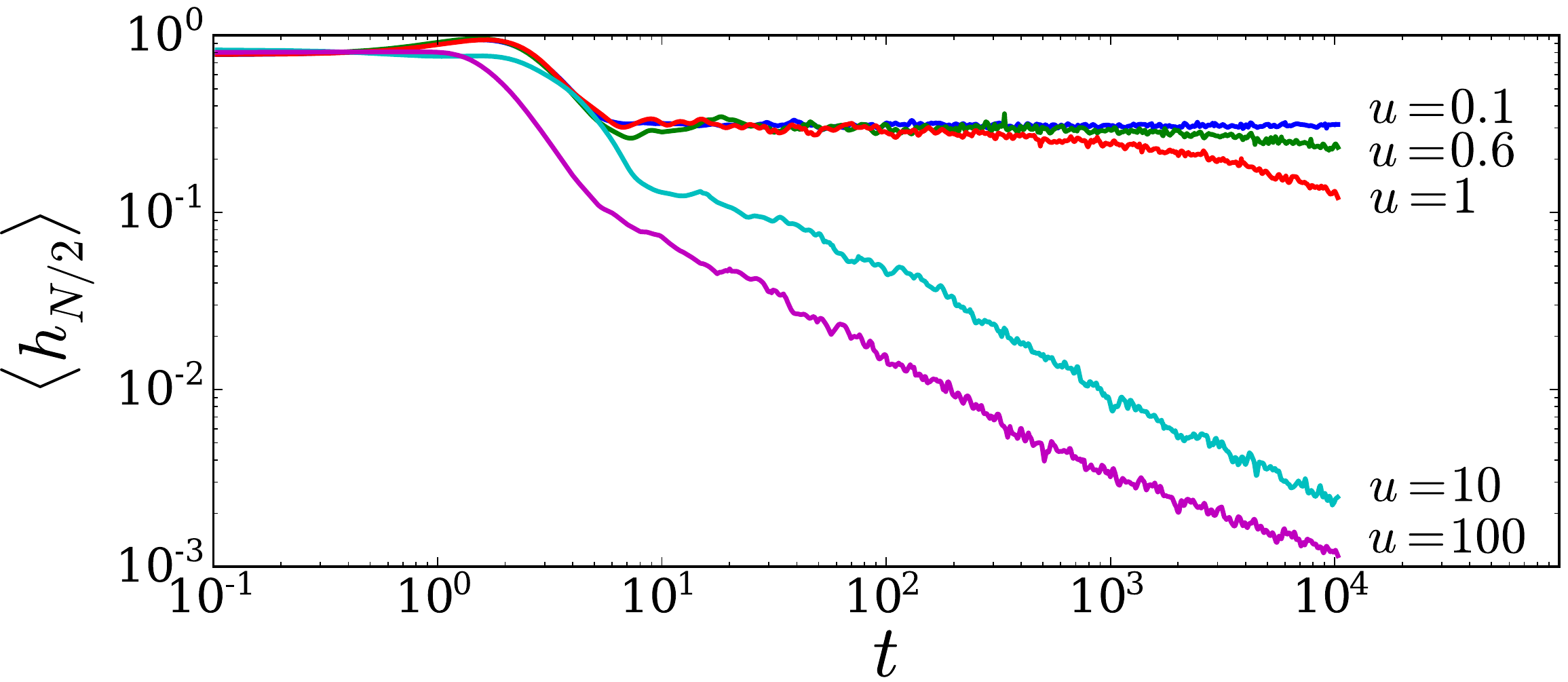}
\caption{(Color online) The normalized mean energy  $\langle h_{N/2}\rangle$ of the central bead at $n=N/2$ as a function of time, for different initial displacements;{ from bottom to top, $u=0.1, 0.6, 1, 10, 100$.}}
\label{central}
\end{figure}

It is also relevant to discuss the behavior of the mean participation number $\langle P\rangle$ in this weakly nonlinear regime. As it is shown in Fig.~\ref{intermediate} (b) after an initial increase of $\langle P \rangle$ its value 
remains practically constant with a value of around {$\langle P \rangle\approx 10$},  between  {$t\approx 10^2$ and $t\approx10^3$}, but after this time interval $\langle P\rangle$ increases  exhibiting a diverging trend. 
This transition is attributed to nonlinearity, since it is never observed in the
near linear regime or in the exact linear case. To further understand the origin of this behavior, we plot in Fig.~\ref{central} 
the normalized mean  energy of the central bead $\langle h_{N/2}\rangle$ as a function of time. 
For $u=0.1$ { after a small transient time of $t\approx 10$, the central bead retains a large amount
of the total energy of the system, keeping it 
up to the end of our simulations at $t=10^4$. }This
behavior is understood by the fact that most of the energy is concentrated around the initially excited bead, due to the presence of the
disorder-induced localized modes. These modes remain localized throughout the simulation.
{ However,  for larger values of the initial excitation, i.e. $0.1< u \lesssim 1$, we observe that although for a 
large time interval ($10\lesssim t \lesssim 10^3$) most of the energy is trapped around the central bead, after sufficient time 
the energy of this bead starts to decrease. This signals the detrapping of energy from this bead, and its release to the rest of the chain.} 


\subsection{Highly nonlinear regime}

\begin{figure}[tb]
\includegraphics[scale=0.35]{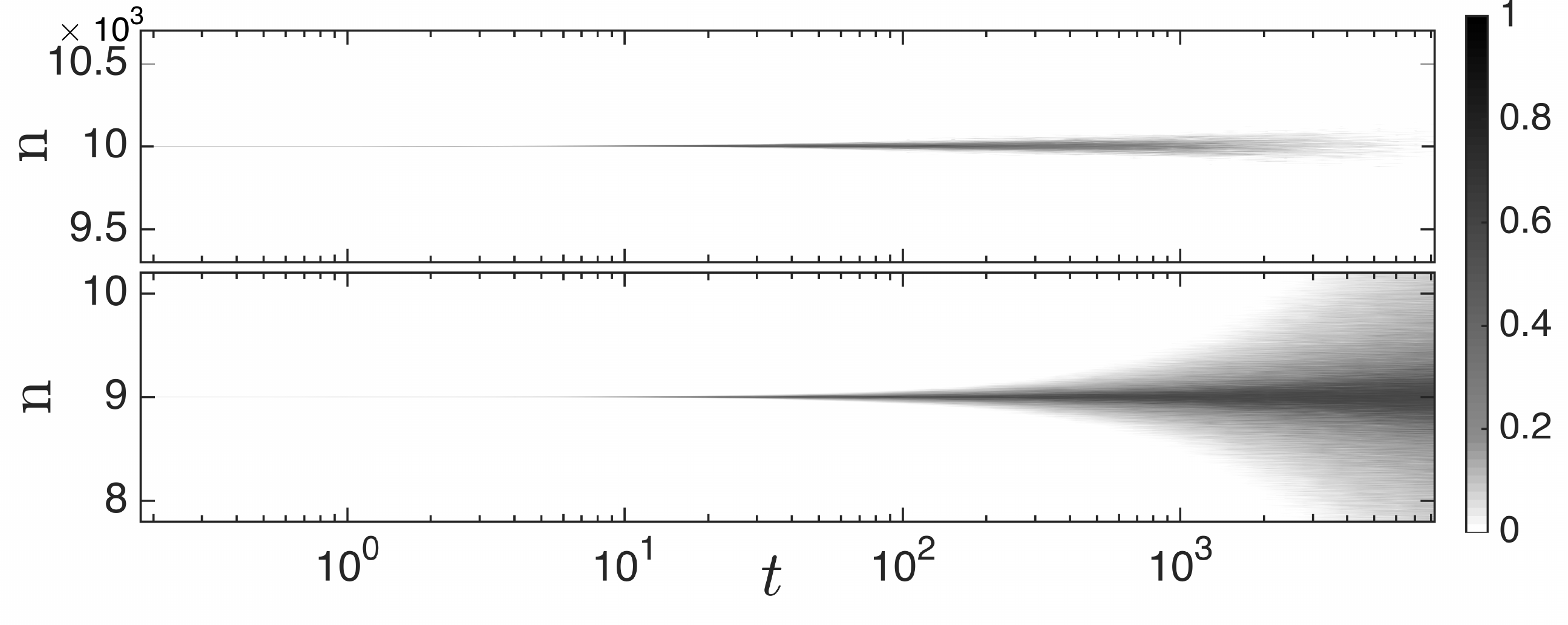}
\caption{The probability of a gap opening as a function of time
for two different initial displacements in the highly nonlinear regime: { $u=10$ (upper panel) and $u=100$ (lower panel)}. We focus on the dynamics around the central bead $N/2$. The black color in the colormap corresponds to probability $1$ while the white to $0$.}
\label{singsiteg}
\end{figure}

For even larger values of the initial condition, $u>1$, i.e. for larger nonlinearities, we observe in Fig.~\ref{singsite}
that the exponent $\beta$ saturates to an almost constant value at about {$t\approx 10^2$} 
describing  a super-diffusive regime. These values are larger with respect to the values of $\beta$ seen in the
weakly nonlinear regime. It is worth noting that in this regime the fluctuations in the values of $\beta$ are much smaller than in 
the previous regime. From Fig.~\ref{central}, we  also conclude that the normalized mean energy of the central bead $\langle h_{N/2}\rangle$ for $u=10,100$ continuously decreases as a function of time, in contrast with the weakly nonlinear regime.
To further investigate the dynamics in this regime we evaluate the probability 
of gap openings between beads as obtained by counting the number of  gaps in each site for all the  200 disorder realizations,
and plot in Fig.~\ref{singsiteg} the obtained average value. In great contrast to the weakly nonlinear regime where no gaps appear, here we find that not only there are always many gaps around the initially excited bead,
but also these gaps propagate in the system. This observation strongly suggests that 
for such large initial excitations, a new dynamical regime is present,  where the dynamics is not governed by the FPU like nonlinearity but by the nonsmooth nonlinearity of the opening of gaps.  We call this regime \textit{highly nonlinear}.

\begin{figure}[tb]
\includegraphics[scale=0.4,trim={80pt 0pt 45pt 20pt},clip]{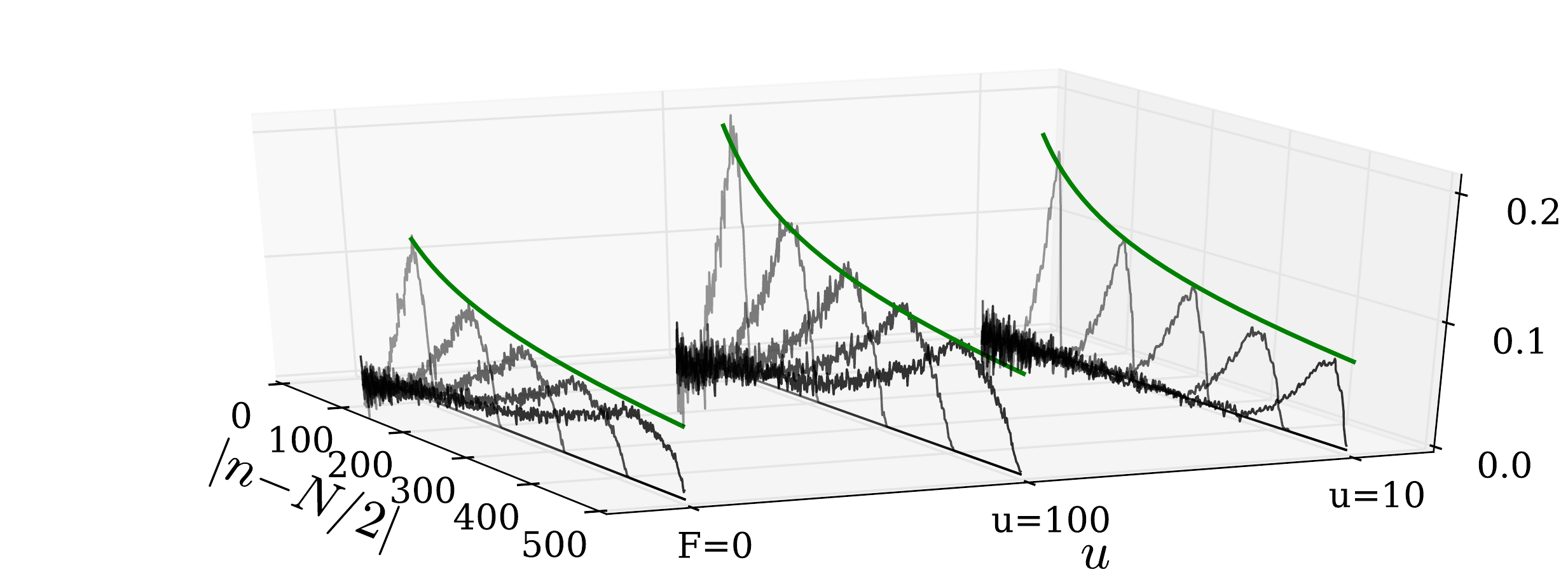}
\put(-55,70){\Large $\frac{\dot{u}_n}{max(\dot{u}_n)}$}
\caption{(Color online) The velocity profiles $\dot{u}_n$ for five time instants, normalized to its maximum value for different initial conditions. Left: the case of $F=0$ with $u=1$, middle and right: the case of  $F\ne 0$ with $u=100$ and $u=10$ respectively.}
\label{singsite2}
\end{figure}

A particular limit of this highly nonlinear regime corresponds to the case of {$F=0$ which results to $\delta_n=0$ in Eq.(\ref{nlineareqm}}). This is also called {\it sonic vacuum}~\cite{compacton} due to the 
fact that the system does not support 
the propagation of linear waves.
This regime has been studied extensively for the case of no disorder, namely when $\alpha=1$. 
It is known that a solitary solution exists in this limit { with a highly localized waveform}~\cite{compacton}. 
For the case of a binary system with a disordered distribution between beads of two different masses, similar solitary
waves of decreasing amplitude were found in the weak disorder limit, while in the strong disorder case a delocalized
wave was observed~\cite{chiaropanos}. Similar results were obtained for the case of two-dimensional granular solids, with an initial excitation only in
one direction: weak disorder induces an exponentially decreasing solitary wave which eventually gives its place to a delocalized shock-like 
profile, while strong disorder only exhibits the shock-like structure~\cite{vitelli1,shock}.
The latter works also showed that at the position  $n_f$ of the front of the shock-like structure, the velocity scales 
as{ $\dot{u}_{ f}\propto n_{f}^{-1/2}$.} 

Let us now explore the transition from the highly nonlinear regime to the singular case of $F=0$.
First note that, as it is shown in Fig.~\ref{singsite} (b), {for $F=0$ the exponent $\beta$ reaches the value
$\beta \approx 1.8$, which is the maximum value}, while the mean participation number is qualitatively similar for all the cases in the highly nonlinear regime. 
Furthermore in Fig.~\ref{singsite2}, we plot the velocity profiles at five time instants for the case of $F=0$ with $u=1$, and for the case of $F\ne 0$ with $u=10$ and $u=100$. 
For $F=0$, in agreement with ~\cite{vitelli1,shock}, we observe the formation of a shock-like structure, {see left case in Fig.~\ref{singsite2}}. In fact, by plotting a fitting curve of the form 
$n_{f}^{-1/2}$, shown as an envelope on-top of the velocity profiles [solid (green) line], it is clearly seen that the propagating front follows this $-1/2$ power law trend.

More importantly we find that even in the case of a finite precompression force, $F\ne 0$, a similar structure can be formed and is propagating with the same power law decay of its amplitude, { see middle case in Fig.~\ref{singsite2}}. 
However, for  the case  of $u=10$ although the peak of the propagating form seems to follow a similar trend,
the observed structure is different; instead of a shock-like profile it appears to be bell-shaped and it consists 
of a smaller number of beads {($\approx$ 100 beads), see right case in Fig.~\ref{singsite2}}. 

\section{Momentum excitation}
\label{s4}

\begin{figure}[tb]
\includegraphics[scale=0.36]{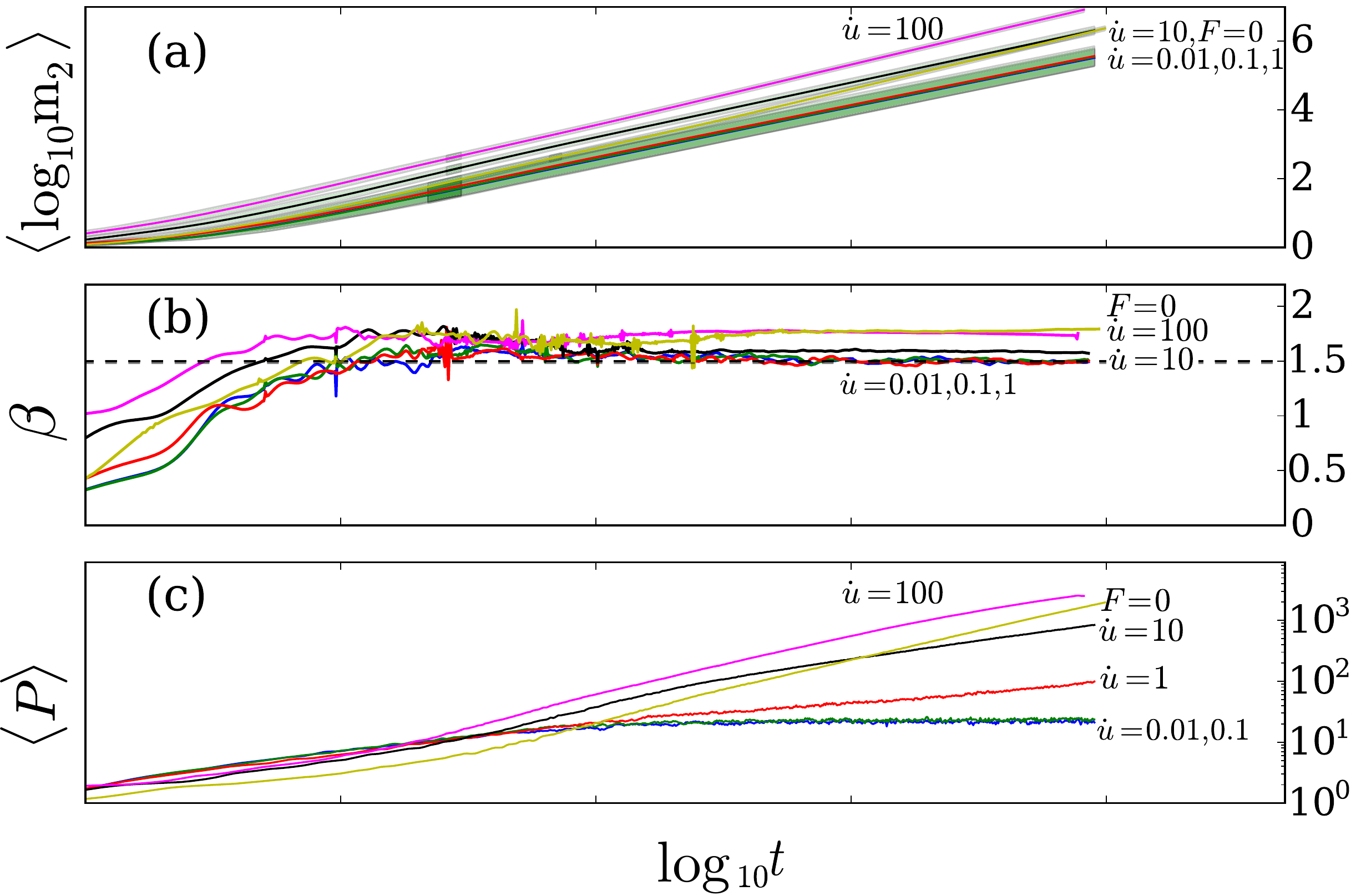}
\caption{(Color online) Same as in Fig.~\ref{singsite} but for different initial momentum excitations $\dot{u}_{N/2}(0)=\dot{u}$.}
\label{impulse}
\end{figure}

\begin{figure}[tb]
\includegraphics[scale=0.37]{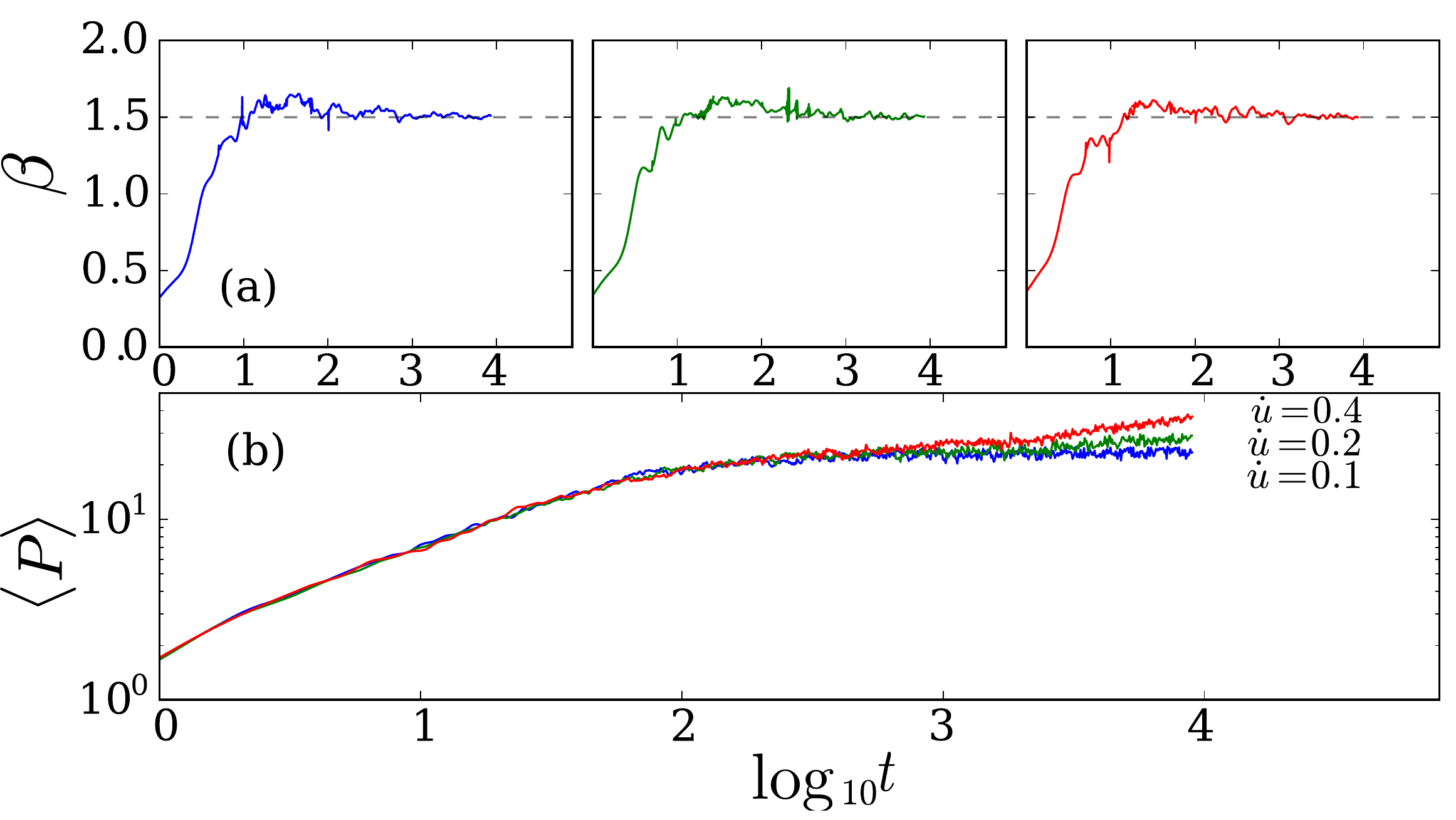}
\caption{[Color online] Same as in Fig.~\ref{intermediate} but for different initial momentum excitations $\dot{u}_{N/2}(0)=\dot{u}$. The top panels correspond to $\dot{u}=0.1, 0.2, 0.4$	from left to right.}
\label{participation2}
\end{figure}

{ As we mentioned in the introduction, the energy transport in disordered linear chains, strongly depends on the initial
condition~\cite{lepriPRE,kundu}.}
Thus in order to complete the dynamical study of our model we also investigate the case of an initial momentum excitation of the central spherical bead by increasing the initial velocity $\dot{u}_{N/2}(0)=\dot{u}$. The results are summarized in Fig.~\ref{impulse} and Fig.~\ref{participation2}. 

In the linear case the energy spreading for an initial momentum excitation is known to be super-diffusive with $\beta=1.5$~\cite{kundu}.
{By numerical simulations 
of the normalized} nonlinear equations of motion~(\ref{nlineareqm}), we found that the energy spreading remains super-diffusive with $\beta=1.5$, for all initial velocities $\dot{u}<10$
[see Fig.~\ref{impulse} (b), Fig.~\ref{participation2} (a)]. This is in contrast with respect to the case of the initial displacement excitation, where the relevant values of $\beta$ exhibit large variations.
For $\dot{u}\gtrsim10$ the exponent $\beta$  grows and eventually reaches the maximum value of  $\beta\sim 1.8$, which is also the corresponding value of $\beta$ of the limiting case of $F=0$. 
We  again identify this regime as the highly nonlinear regime.

On the other hand, the mean participation number $\langle P\rangle$
exhibits similar behavior with the case of initial displacement excitations. For $\dot{u}\lesssim 0.1$ it
saturates to a constant value of about $20$ beads, while for $\dot{u}>0.1$ it {continuously increases}. 
Looking closer to $\langle P\rangle$, shown in Fig.~\ref{participation2}(b) 
we note that the mean participation number reaches the asymptotic constant value of about $20$ beads for $\dot{u}=0.1$, i.e., the near linear
regime. However, for $\dot{u}=0.2,0.4$ although it seems to saturate to a constant value for long time, finally it starts
to deviate and to {increase continuously}. This again suggests that there is an intermediate regime for $0.1< \dot{u} 
\lesssim 1$
{ in which energy detrapping is observed}. This is the weakly nonlinear regime.

To conclude, compared to the case of displacement initial excitations, there is a difference in the energy transport properties in the intermediate regime that we call weakly nonlinear. Although the mean participation number  $\langle P\rangle$ shows the same behavior, in contrast to the displacement excitations, for momentum initial excitations both the near linear and weakly nonlinear regimes are characterized by an asymptotic value of the parameter $\beta=1.5$, namely 
the same as to the linear case. 

Another interesting point to be mentioned is {that} although in the linear case  the energy transfer is 
sub-diffusive or super-diffusive for an initial displacement or an initial momentum excitation respectively, in the 
highly nonlinear regime and more profoundly in the limiting case
of $F=0$ both excitations result in the same behavior of energy transport. { In fact in this limit, not only 
the asymptotic value of $\beta$ (as it was also found in Ref.~\cite{Mason2} for a disordered dimer),} but also  the dynamics of the
derivative $\beta$ of $\langle m_2 \rangle$ are very similar.

\section{Asymptotic profiles}
\label{s5}
\begin{figure*}[tb]
\includegraphics[scale=0.45]{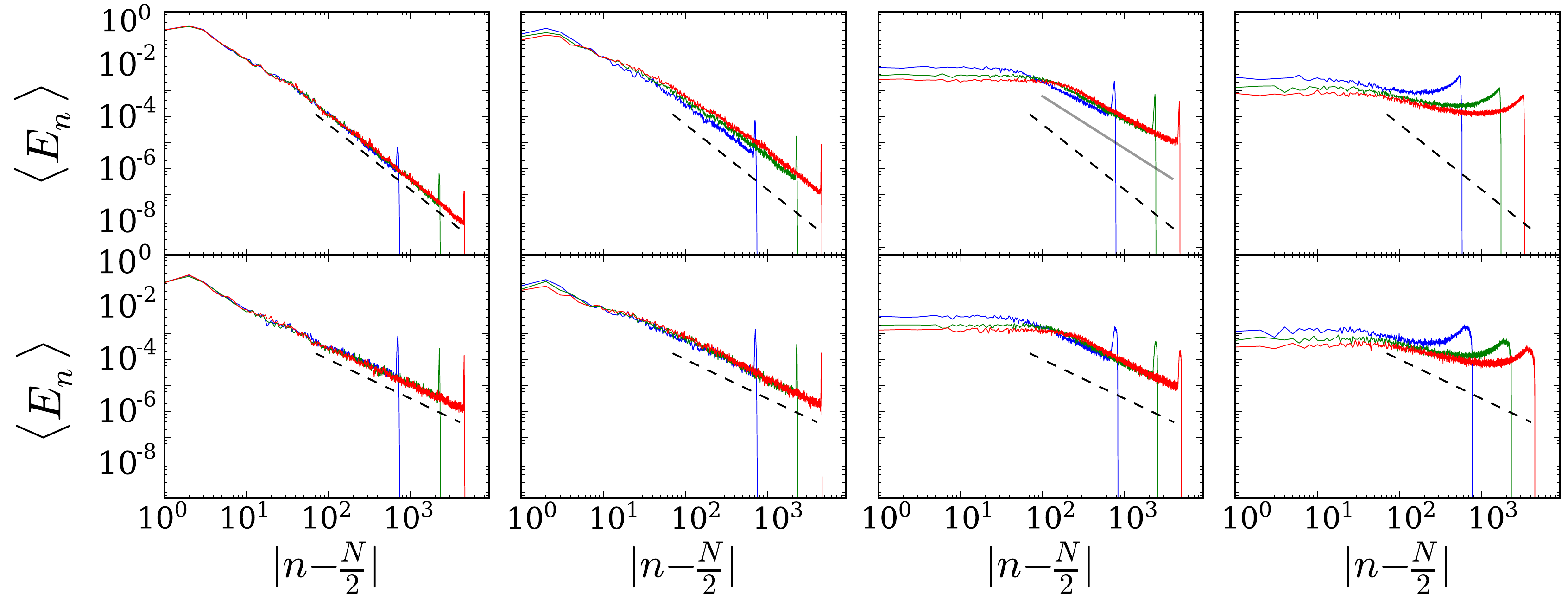}
\caption{(Color online) Profiles of the energy distribution for initial displacement (top panels) and momentum (bottom panels)
excitation. The values of the initial {displacement excitations are $u=0.01, 1, 10$ for the first three top panels (frome left to right) and $u=1$ when $F=0$ for the fourth top panel. The values of the initial momentum excitations are $\dot{u}=0.01, 1, 10$ for the first three top panels (from left to right) and $\dot{u}=1$ when $F=0$ for the fourth top panel.
These values cover all the dynamical regimes: from 
the near linear (left column) to the highly nonlinear (right column) regime. The different profiles are taken at times $t\approx 1500$ (blue), $t\approx 5000$ (green) and $t \approx 10000$ (red). The dashed lines denote slopes of $5/2$ (top) and $3/2$ (bottom). For the top third panel, the additional solid line denotes the slope of $3/2$}.}
\label{lepri}
\end{figure*}

According to the recent work of Ref.~\cite{lepriPRE},  the asymptotic
dependence of the energy moments (not only $m_2$) can be characterized by the asymptotic energy profile
of the lattice, at least in the linear case. In particular, it was shown in that work, that the energy profile far away from the central excitation has the 
form of $\langle h_n\rangle\sim |n-N/2|^{-\eta}$. The exponent $\eta$ was found to be
$\eta=5/2$  and $\eta=3/2$  for displacement and momentum initial excitations respectively.
In Fig.~\ref{lepri} we show for both displacement (top panels) and momentum initial excitations (bottom panels) three different instants of the energy profile at sufficiently 
large times, covering all the dynamical regimes from near linear (left column) to highly nonlinear (right column). In the left panels (near linear regime)
it is readily seen that, in accordance with the predictions for the linear problem, the three profiles overlap indicating the fact
that the energy distribution has reached an asymptotic profile which does not change for later times. For comparison, we
also plot  curves (dashed lines) with a slope of $5/2$ (top) and $3/2$ (bottom). Note also that the energy distribution near the central bead
(i.e.~for $n\approx N/2$ in the figure) is the same for the three different profiles. This indicates that the energy of these sites also does not change in time, and the mode around the center remains localized.

The top (bottom) second column {panel} of Fig.~\ref{lepri}, depicts the energy profile
for an initial displacement (momentum) excitation with $u=1$ ($\dot{u}=1$). In this case it is readily seen that for the initial displacement excitation (top) not only the three profiles do not overlap, but also the slopes are not
exactly $5/2$. This is another indication for the appearance of the weakly nonlinear regime which exhibits nontrivial dynamics. Note that for the momentum excitation (bottom) the profiles do overlap. { This is in accordance with the discussion in section V about the asymptotic value of $\beta$ which is the same for both the near linear and the weakly nonlinear regime}.
Additionally, for both cases of initial conditions, we observe a small but non negligible deviation of the energy around the center, which confirms
the fact that the localized mode around the center starts to loose its energy.

For larger values of the initial condition as is shown in the top (bottom) third column {panel} of Fig.~\ref{lepri}, the profile of the energy is substantially different.
It is characterized by two different regimes: a weakly localized part with almost $10^2$
beads {with similar amount of energy, forming an almost straight horizontal line} and a decaying tail.
However, the weakly localized portion of the energy is spreading and loses its energy, as for larger times the almost straight horizontal part of the profile becomes longer having smaller energy values. We also note that once again the slope of the energy profiles for initial displacements (top panel) is not  $5/2$ {as time increases} but interestingly enough it reaches a value which is closer to $3/2$ (see gray solid line). 
We remind  that for this initial displacement excitation with $u=10$, $\beta$ asymptotically reaches the value of $1.5$ (see Fig.~\ref{singsite}) which is also the asymptotic value of $\beta$ of an initial momentum excitation of the linear problem.

Finally for the particular case of $F=0$, as shown in the top (bottom) fourth column {panel} of  Fig.~\ref{lepri}, the asymptotic energy profile is similar to a ballistic propagation where 
{the energy differences between excited beads decrease drastically}
but in this case the propagating front does not exhibit a sharp profile. In fact this front has a very large ``tail'' which is due to the
shock-like structure 
 {
 that was  mentioned in section C}.


\section{Conclusions}
\label{s6}
In this work we numerically investigated the energy transport in a one-dimensional  granular solid composed of spherical beads of randomly 
distributed radii which interact via Hertzian forces. We studied the dynamics by using two different localized initial conditions 
i.e.~initial displacement and initial velocity excitations of the central bead of the chain and by increasing the amplitude of these excitations.
We were able to identify three different dynamical spreading regimes with distinct characteristics: the near linear, the weakly nonlinear and the highly nonlinear.

In the near linear regime, part of the initial energy remains localized around the central excited bead while
two counter-propagating fronts, coherent phonons, travel through the chain. 
We found that the energy transport is identical to that of a linear chain, with either mass or coupling 
disorder, at least up to the time scales reached in our simulations. The spreading of the energy 
is characterized  by an asymptotic time dependence of the mean second moment of the energy  $m_2$  of the form 
$\langle m_2\rangle\sim t^\beta$, where  $\beta=0.5$ and $\beta=1.5$ for an initial displacement and an initial
momentum excitation respectively.  Additionally, the mean participation number $\langle P\rangle$ in this regime,
{ converges to a constant value}, which depends on the strength of the disorder.
 
For larger values of the initial conditions, in the weakly nonlinear regime, we found that for initial displacement excitations the energy spreading
does not exhibit a clear asymptotic time dependence. However it does cross to a super-diffusive behavior,
since for large enough timescales, the slope of $m_2$ becomes larger than 1. 
In fact this behavior, which was also observed in the recent study of~\cite{lepriPRE} for an FPU lattice,
is found to be closely connected with the nonlinear dynamics of the localized state formed around the central spherical bead.
We found that after a sufficient time interval (between $10^2$ and $10^3$ { normalized time units}; see Eq.(\ref{norm})), the central 
localized  region consisting of about $10$ beads, starts to delocalize and the energy stored in these spherical beads starts to 
radiate into the system. In this weakly nonlinear regime, the  dynamics is governed by the power Hertzian nonlinearity. 
On the other hand, for initial momentum excitations, the slope of $m_2$ remains around $1.5$, as in the near linear regime, but
 $\langle P\rangle$ exhibits the same behavior as for displacement excitations. 

For even larger amplitudes of the initial excitation, the energy transfer becomes substantially different. { The energy profile
of the chain reveals an almost ballistic behavior with an almost equal distribution of the energy around the excited portion of the chain}.
{In this regime, which we characterized highly nonlinear, the system exhibits a large number of opening of gaps between beads, which is a nonsmooth nonlinear process.}  
We found that these gaps propagate in the chain, and  that the transport of energy is mediated
by a shock-like structure,  which bares similarities with the selfsimilar solution found in~\cite{shock}.
 
An important result of our work is the following: {although it is known that the energy transport for the disordered \textit{linear} chain strongly depends
on the type of initial conditions (i.e.  displacement or momentum excitations), we found that in the \textit{highly nonlinear} regime it is
independent of  the initial condition. This is a rather general feature of disordered granular chains as it was very recently observed in different disordered dimer granular chain~\cite{Mason2}}. In particular, in the linear case the asymptotic time dependence of $\langle m_2\rangle$ shows
a slope of $\beta=0.5$ (displacement) and $\beta=1.5$ (momentum), while for the extreme nonlinear limit
of $F=0$ both initial conditions lead to a slope of $\beta\approx 1.8$. Additionally, the energy profiles in this regime
show that there is no distinct localized state. 

\section*{Acknowledgments}
G.~T. acknowledges financial support from FP7-CIG (Project 618322 ComGranSol).
Ch.~S. was partially supported by the National Research Foundation of
South Africa {(Incentive Funding for Rated Researchers, IFRR). We thank the anonymous referees whose remarks helped us improve the clarity of the paper. Ch.S thanks LAUM for its hospitality during his visits in June and November 2015, when part of this work was carried out}.

\end{document}